\documentclass[preprint,preprintnumbers,amsmath,amssymb]{revtex4}
\usepackage{amsfonts}    
\usepackage{amssymb}
\usepackage{latexsym}
\usepackage{eepic}
\usepackage{graphicx}
\usepackage[usenames,dvipsnames]{color}

\usepackage[active]{srcltx}

\newcommand{\al}{\alpha}

\newcommand{\De}{\Delta}

\newcommand{\rar}{\rightarrow}
\newcommand{\lrar}{\leftrightarrow}

\begin{document}

\title{Helium- and Lithium-like ionic sequences: Critical charges}

\author{N.L.~Guevara}
\email{nicolais@phys.ksu.edu}
\affiliation{
Department of Physics, Kansas State University, Manhattan, KS 66506, USA}

\author{A.V.~Turbiner}
\email{turbiner@nucleares.unam.mx}
\affiliation{Instituto de Ciencias Nucleares, Universidad Nacional
Aut\'onoma de M\'exico, Apartado Postal 70-543, 04510 M\'exico,
D.F., Mexico}

\begin{abstract}
In non-relativistic quantum mechanics we study the Coulomb systems of infinitely massive center of charge Z and two-three electrons: $(Z,e,e)$ and $(Z,e,e,e)$. It is shown that
in both cases the total energy curve in $Z$ is smooth, without any visible irregularities. Thus, for both systems the physical integer charges $Z=1,2,\ldots$ do not play a distinguished role as would be associated with charge quantization. By definition, a critical charge $Z_{cr}$ is a charge which separates a domain of the existence of bound states from a domain of unbound ones (continuum). For both systems the critical charges are found, $Z_{cr,2e}=0.91085$ and $Z_{cr,3e}=2.009$, respectively. Based on numerical analysis, the Puiseux expansion in fractional powers of $(Z-Z_{cr})$ is constructed for both systems. Our results indicate the existence of a square-root branch point singularity at $Z_{cr}$ with exponent 3/2. A connection between the critical charge and the radius of convergence of $1/Z$-expansion is briefly discussed.

\end{abstract}

\pacs{31.15.Pf,31.10.+z,32.60.+i,97.10.Ld}

\maketitle

Let us consider the Hamiltonian which describes the charged center $Z$ and $k$ electrons
\begin{equation}
\label{H}
    {\cal H}\ =\ -\frac{1}{2} \sum_{a=1}^k \De_a \ -\
    \sum_{a=1}^{k} \frac{Z}{r_{a}}  \ +\
    \sum^k_{a<b} \frac{1}{r_{ab}}\ .
\end{equation}
The case $Z=k$ corresponds to a neutral atom and $Z=k+1,2,\ldots$ describes positive ions.
Sometimes, there exist negative ions $Z<k$, e.g. at $Z=1,k=2$.
Critical charge $Z_{cr}$ is a value of $Z$ which separates the domains "existence/non-existence" of square-integrable solution(s) of the Schroedinger equation,
${\cal H} \Psi= E(Z) \Psi$. In other words, it separates the domain, of charge $Z$, of the Hamiltonian where the bound state(s) exists from the domain where it does not. Constructively, $Z_{cr}$ is determined by a condition of vanishing the ionization energy.
The goal of this short Note is to find the critical charge $Z_{cr}$ for two electron (Helium-like) and for three-electron (Lithium-like) sequence. It was claimed long ago that the question about critical charge is closely related to one of the first theoretical questions of newly born atomic physics about the radius of convergence $Z_*$ in the $1/Z$-expansion. Making change of variables in (\ref{H}), $\vec{r} \rar \vec{r}/Z$, we arrive at
\begin{equation}
\label{H_t}
    {\cal H}_t\ =\ -\frac{1}{2} \sum_{a=1}^k \De_a
    \ -\ \sum_{a=1}^{k} \frac{1}{r_{a}}  \ +\ \frac{1}{Z}\sum^k_{a<b} \frac{1}{r_{ab}}
\end{equation}
where the new energy $\tilde E(\xi=\frac{1}{Z})=\frac{E(Z)}{Z^2}$. In general, for all studied cases the ground state energy $\tilde E(\xi)$ is a smooth, slow-changing real function in $\xi \in [0, \xi_{cr}]$.
Since time of Hylleraas it was said that for two-electron case, $k=2$, the radius of convergence $\xi_*$ for the expansion of $\tilde E$,
\begin{equation}
\label{E_t}
    \tilde E\ =\ \sum_{n=0}^{\infty} e_n \xi^n
\end{equation}
coincides with $\xi_*=1/Z_{cr}$. Furthermore, it was claimed that the nearest singularity to $\xi=0$, which defines the radius of convergence, is situated at real $\xi$-axis and even at $\xi=\xi_*=\xi_{cr}\equiv 1/Z_{cr}$ (see \cite{Baker:1990} and references therein where extensive discussion with extended bibliography together with historical account are presented). Based on the analysis of large number of computed coefficients of $1/Z$-expansion $(\sim 401)$ it was found that the critical charge for two-electron sequence is
\begin{equation}
\label{ZcrHe1}
    Z_{cr,2e}^{[1]}\ =\ 0.91103... \ ,
\end{equation}
and the singularity associated with it is an essential singularity of quite complicated nature \cite{Baker:1990}. Although this value of the critical charge is actually quite close to the first estimate of the critical charge given by Stillinger and Stillinger  \cite{Stillinger:1974}
\begin{equation}
\label{ZcrHe2}
    Z_{cr,2e}^{[2]}\ =\ 0.8941... \ ,
\end{equation}
it is in contradiction with the nature of the singularity. It is claimed in \cite{Stillinger:1974} (see Eq. (2.17)) that it is a branch point with exponent 3/2.
Recently, the result (\ref{ZcrHe1}) was challenged in \cite{Zamastil:2010} by using the accurate numerical analysis of a different set of coefficients in $1/Z$-expansion than one used in \cite{Baker:1990}. It is stated in \cite{Zamastil:2010} that
\begin{equation}
\label{ZcrHe3}
    Z_{cr,2e}^{[3]}\ =\ 0.90223... \ .
\end{equation}
This result agrees with \cite{Stillinger:1974}  on the nature of the singularity as a branch point with exponent 3/2. Our analysis unrelated to perturbation theory excludes the results (\ref{ZcrHe2})-(\ref{ZcrHe3}) about a value of a critical charge as well as about a position of the singularity defining the radius of convergence of (\ref{E_t}). Our critical charge is close to $Z_{cr}^{[1]}$ (see below). However, we confirm the observation by \cite{Stillinger:1974}, \cite{Zamastil:2010} that the singularity at the critical charge is a branch point with exponent 3/2. We are not certain about a nature of singularity(ies) which define the radius of convergence of the $1/Z-$expansion (\ref{E_t}).

Our approach is not related with an analysis of $1/Z$-expansion being instead based on accurate calculation of the total energy as a function of the charge $Z$ in vicinity of the critical point with further extrapolation to the critical point. Extrapolating function is assumed to be in a form of the Puiseux expansion for $Z > Z_{cr}$,
\begin{equation}
\label{Puiseux}
    E(Z)\ =\ \sum_{n=0}^{\infty} B_n (Z - Z_{cr})^{\al_n}\ ,\
\end{equation}
with a condition that exponents $\al_{n}$ should grow with the increase of $n$, $\al_{n} < \al_{n+1}$. Both cases of two and three-electron systems are considered.

{\bf Helium-like sequence.} In order to find the total energy of the $1^1S$ state (the ground state) as the function of $Z$ we used as trial function a linear superposition of the exponential, explicitly correlated functions (see e.g. \cite{Korobov:2000})
\begin{equation}
\label{trial2}
  \Psi_{trial}=\sum_{i=0}^N A_i \left[\exp{(-\al_i r_1 -\beta_i r_2)}+(1 \lrar 2)\right]\exp{(-a_i r_{12})} \ ,
\end{equation}
where $\{A_i\}$ and $\{ \al_i, \beta_i, a_i \}$ are linear and non-linear parameters, respectively. It was shown that this basis provides at present the fastest convergence in ground state energy for Helium $(Z=2)$ and H$^-$ $(Z=1)$ among known bases. Namely using this basis the ground state energy for $Z=1,2$ was found with 24 s.d. \cite{Korobov:2000}.

Using the computer code kindly provided by Prof. K.~Pachucki (see www.fuw.edu.pl/~krp/) designed following the work \cite{Korobov:2000} we see that the fast convergence in energy is also obtained for $Z$
other than 1 and 2. We calculated the ground state energies for different $Z$ varying from 0.95 up to 2.00 with interval 0.05 with 12 significant digits (see Table I for some examples). It is worth noting that for $Z < 0.95$ the rate of convergence starts to deteriorate dramatically with a decrease of $Z$ - it requires many more terms in (\ref{trial2}) to be included to reach desired accuracy unlike for $Z > 0.95$. We exclude the domain $Z < 0.95$ from calculation of energies based on (\ref{trial2}).

Calculations of the ground state energy for different values of $\xi$ (in 30 points) confirm that the energy $\tilde E_{2e}(\xi)$ is a smooth, slow-changing function without any visible irregularities in domain $[0, 1/Z_{cr,2e}]$. It does not exhibit any distinguished role of points $1/M, M = 1,2,\ldots$, which would indicate the appearance
of the charge quantization.

\begin{table}[h]
  \centering
\caption{
\label{table1}
 Ground state energy $E_{2e}$ for two-electron sequence for selected values of Z found using (\ref{trial2}) where all digits assumed to be correct and $E_{2e}^{(fit)}$ from the fit (\ref{e-fit_2}).
}
\begin{tabular}{|c|c|c|}
\hline\hline
Z          &\       $E_{2e}$        \ &\  $E_{2e}^{(fit)}$\  \\
\tableline
\ 1.3   \  &\  -1.029896662309   \ &\  -1.029896664 \  \\
\ 1.25  \  &\  -0.933575272295   \ &\  -0.933575273 \  \\
\ 1.15  \  &\  -0.756014315641   \ &\  -0.756014316 \  \\
\ 0.95  \  &\  -0.462124684391   \ &\  -0.462124684 \  \\
\tableline
\hline
\end{tabular}
\end{table}

We make fit with terminated expansion (\ref{Puiseux}) whose length is subsequently increased from $n=3$ up to $n=7$. Taking a large number of points in the interval $Z \in [0.95, 1.5]$ and making fit it is obtained in a stable way with sufficiently high accuracy that $\al_n$ takes either integer or half-integer values(!),
\begin{equation}
\label{alfas}
 \al_0=0\ ,\ \al_2=1\ , \al_3=3/2\ ,\ \al_4=2\ ,\ \al_5=5/2\ ,\ \al_6=3\ ,\ \al_7=7/2 \ .
\end{equation}
It confirms, in particular, the rigorous result by Simon \cite{Simon:1977} about the absence of the square-root term, $\al=1/2$ and presence of the linear term $\al=1$ in this expansion. It contradicts a statement from \cite{Kais:1997} that the term $\al=3/2$ is absent. In general, such an expansion
is in agreement with one proposed in \cite{Stillinger:1974, Zamastil:2010}. Eventually, assuming the exponents (\ref{alfas}) and taking data set for energy $E_{2e}$ at nine points $Z \in [0.95, 1., 1.05, 1.1, 1.15, 1.2, 1.25, 1.3, 1.35]$  a careful interpolation leads to the expansion (see (\ref{Puiseux})),
\[
 E_{2e}^{(fit)}(Z)\ =\ -\frac{Z_{cr}^2}{2} - 1.142552  (Z - Z_{cr})
 - 0.174110 (Z - Z_{cr})^{3/2} - 0.7700097 (Z - Z_{cr})^2\
\]
\begin{equation}
\label{e-fit_2}
    - 0.1399230 (Z - Z_{cr})^{5/2} + 0.0224694 (Z - Z_{cr})^3
    + 0.0087298 (Z - Z_{cr})^{7/2} \ ,
\end{equation}
with the critical charge
\begin{equation}
\label{Zcr2e}
    Z_{cr,2e}\ =\ 0.91085\ .
\end{equation}
It can be seen that the expression (\ref{e-fit_2}) reproduces 8-9 s.d. in all numbers for energies included in Table~I.

The explicit knowledge of the first terms of the Puiseux expansion (\ref{e-fit_2}) allows us to check whether the singularity at the critical charge defines the radius of convergence of the expansion (\ref{E_t}). In order to do it we construct the Puiseux expansion of the function $\tilde E_{2e}(\xi)$ near the critical $\xi_{cr}=1/Z_{cr}$,
\[
    \tilde E_{2e} (\xi) \ =\ -\frac{1}{2} + \tilde  B_1 (\xi_{cr}-\xi) + \frac{B_3}{\xi_{cr}} (\xi_{cr}-\xi)^{3/2} +  \tilde  B_4 (\xi_{cr}-\xi)^2 + \frac{1}{\xi_{cr}^3}\left(B_5 - \xi_{cr}\frac{B_3}{2}\right) (\xi_{cr}-\xi)^{5/2} +
\]
\begin{equation}
\label{e-fit_2-xi}
    \frac{B_6 }{\xi_c^4}(\xi_{cr}-\xi)^3 +
    \frac{1}{\xi_{cr}^5}\left(B_7 + \xi_{cr}\frac{B_5}{2} - \xi_{cr}^2\frac{B_3}{8}\right) (\xi_{cr}-\xi)^{7/2} + \ldots\ ,
\end{equation}
where $\tilde B_{0,1,4}$ are related with coefficients in the expansion (\ref{Puiseux}), (\ref{e-fit_2}). It leads to the following form of the coefficient $e_n$
in (\ref{E_t}) at large $n$,
\[
   e_n = (-)^n \frac{\Gamma (\frac{5}{2})}{\Gamma (n+1) \Gamma (\frac{5}{2}-n)}\ \frac{1}{\xi_{cr}^{n-\frac{1}{2}}}
\]
\begin{equation}
\label{alpha}
\bigg[B_3  + \left(B_5 - \xi_{cr}\frac{B_3}{2}\right) \frac{5}{5-2n} \xi_{cr}^{-1} +
   \left(B_7 + \xi_{cr}\frac{B_5}{2} - \xi_{cr}^2\frac{B_3}{8}\right) \frac{35}{(5-2n)(7-2n)}\xi_{cr}^{-2} + \ldots\bigg]\ ,
\end{equation}
which is a type of $1/n-$expansion for $e_n$.
Now we can make a comparison of these coefficients with ones calculated in \cite{Baker:1990}, see Table~\ref{table2}. One can see that for $n > 20$, the coefficients are becoming sufficiently close.
However, a deviation in coefficients is growing with an increase of $n$. For $n=300$, where $1/n$-corrections to the asymptotic behavior of $e_n$ at $n$ tending to infinity presumably can be neglected, the deviation reaches $\sim$6 times when the coefficients are more or less of the same order of magnitude $10^{-20}$. If, for a moment, we exclude a possibility that the large-order coefficients in \cite{Baker:1990} might be calculated incorrectly (see e.g. \cite{Zamastil:2010}) we arrive at the immediate conclusion that the singularity related with the critical charge (\ref{Zcr2e}) defined by the Puiseux expansion (\ref{e-fit_2-xi}) can not explain the large-$n$ behavior of the $e-$coefficients. Hence, there must exist other singularity(ies) on the circle of convergence which define the asymptotic behavior of $e_n$ in (\ref{E_t}) at large $n$.

\begin{table}[h]
  \centering
\caption{\label{table2} Comparison the $\al_n$-coefficients in the expansion (\ref{E_t}) calculated in \cite{Baker:1990} (rounded to 3-5 s.d.), obtained in \cite{Zamastil:2010} and found by using the formula (\ref{alpha}).
}
\begin{tabular}{|c|c|c|c|}
\hline\hline
  n       & \cite{Baker:1990} & \cite{Zamastil:2010} & (\ref{alpha}) \\
\tableline
 20       &  -0.76862e-5      &   -0.76862e-5  & -0.71492e-5  \\
100       &  -0.398e-10       &   --           & -0.689e-10   \\
200       &  -0.301e-15       &   -0.222e-15   & -1.065e-15   \\
300       &  -0.522e-20       &   --           & -3.396e-20   \\
\tableline\hline
\end{tabular}
\end{table}

Indeed, such singularities might naturally exist. Certainly, there exist
the square-root branch points, which appear in Landau-Zener theory of level
quasi-crossings (see e.g. \cite{LL}) due to the level crossing (for discussion
see e.g. the case of quartic anharmonic  oscillator \cite{BW:1969, EG:2009}).
The most natural candidate for such a quasi-crossing might be a pair of two complex-conjugated square-root branch points due to the crossing of spin-singlet
$1 {}^1 S$ and $2 {}^1 S$ states, which have the closest energies at real $Z$ when
both states coexist. Following this idea we assume that the behavior of the energy is
\[
  \tilde E_{2e} (\xi) \ =\ \sqrt{(\xi+a)^2+b^2} \left( A_1 + A_2 [(\xi+a)^2+b^2] +
  \ldots \right)
\]
near square-root branch points at $a \pm \imath b$. We were unable to find $a,b, A_1, A_2$ with $r=(a^2+b^2)^{1/2} < \frac{1}{Z_{cr,2e}}$ which would reproduce the behavior of $e_n$ coefficients at $n>100$ found in \cite{Baker:1990}. It seems that the only possibility which is left is to assume the existence of three (or more) singularities on the circle of convergence: one of them is of the critical charge and the others are pair(s) due to level crossing(s). This possibility looks quite exotic and thus unlikely. Before going to
explore it we think that the independent calculation of the large-order coefficients $e_n$ in (\ref{E_t}) is highly needed (see below {\it Note Added}).

{\bf Lithium-like sequence.} In order to find the ground state energy of $(Z,3e)$-system as the function of $Z$ we use the variational methods with Hylleraas basis set as a trial function \cite{Drake:1995, Pachucki:2006}. The ground-state wave function $\Psi$ is expressed as a linear combination of $\psi$, the antisymmetrized product of $\phi$ and the spin function $\chi$,
\[
   \psi\ =\ {\cal A}[\phi (\vec{r_1},\vec{r_2},\vec{r_3})\chi]
\]
\begin{equation}
\label{trial3}
  \phi_{trial}\ =\ \left[r_1^{n_1} r_2^{n_2} r_3^{n_3} r_{12}^{n_4} r_{13}^{n_5}r_{23}^{n_6}
  \exp{(-a_i r_1 -b_i r_2-c_i r_3)}\right] \ ,
\end{equation}
\[
   \chi\ =\ \al(1)\beta(2)\al(3) - \beta(1)\al(2) \al(3)
\]
where $\{ a_i, b_i, c_i \}$ are positive real parameters and $n_{1-6}$ are non-negative integers \footnote{The second spin function is not taken into account in this calculation, for discussion see e.g. \cite{Harris:2009}}. Using this basis set with 15 variational (non-linear) parameters plus analytic evaluation of some Hylleraas  integrals as well as recursion relations, Puchalski et al \cite{Pachucki:2006} have reached the impressive accuracy $10^{-14}$ in the ground state energy for both $Z=3$ (Li-atom) and $Z=4$ (the Be${}^+$ - ion) which is the highest at present. The results of these authors also indicate that the accuracy of the previous calculation \cite{Drake:1995} is overestimated and is limited to 10 s.d. Making use the computer code employed in \cite{Pachucki:2006}, and kindly provided by Prof. K.~Pachucki (see /www.fuw.edu.pl/~krp/) and modified by adding MINUIT minimization routine from CERN-LIB we are able to see that the fast convergence in energy is also obtained for $Z$ other than 3 and 4.

Calculations of the ground state energy for different values of $\xi$ confirm that the function $\tilde E_{3e}(\xi)$ is a smooth, slow-changing function without any visible irregularities in domain $[0, 1/Z_{cr,3e}]$ (see below). It does not indicate any distinguished role of points $1/M, M = 3, 4, \ldots$, whose would be associated with the appearance of charge quantization.

We calculated the total energies for different $Z$ varying from 2.02 up to 3.00 and reached 7 s.d. with total basis length equals to 502 \footnote{Stability of these 7 s.d. was checked directly making a control calculation with the (optimized) basis length equals to 918. We observe that the accuracy starts to deteriorate dramatically for $Z < 2.02$. We never used data from this domain for analysis.}. To make fit we assume the same exponents (\ref{alfas}) in the Puiseux expansion. Data set which was used for interpolation contains seven points $Z \in [2.02, 2.03, 2.07, 2.08, 2.10, 2.12, 2.16]$ (see Table III). We arrive at the interpolation in a form of the expansion (\ref{Puiseux})
\begin{equation}
\label{e-fit_3}
 E_{3e}^{(fit)}(Z)\ =\ -2.934278 - 3.390491 (Z - Z_{cr,3e})
 - 0.114813 (Z - Z_{cr,3e})^{3/2} - 1.102097 (Z - Z_{cr,3e})^2\ ,
\end{equation}
(cf. (\ref{e-fit_2})), where the critical charge
\begin{equation}
\label{Zcr3e}
    Z_{cr,3e}\ =\ 2.009\ .
\end{equation}
In order to check consistency we found that (\ref{e-fit_3}) at the critical point $Z=Z_{cr,3e}$ reproduces the ground state energy of 2e-system $E_{2e}$ in 6 s.d. In general, the expression (\ref{e-fit_3}) reproduces 6-7 s.d. in all numbers for energies included in Table~III. It is worth noting that the critical charge (\ref{Zcr3e}) is inside the intervals for critical charges proposed in \cite{Kais:1998}. It seems the method of analysis used in \cite{Kais:1998} are quite rough, leading to not very precise results - a direct calculation of the ionization energy $\propto (E_{3e}(Z) - E_{2e}(Z))$ immediately exclude essential parts of those intervals for critical charge.

\begin{table}[h]
  \centering
\caption{
\label{table3}
Ground state energy $E_{3e}$ for three-electron sequence for selected values of Z found using (\ref{trial3}) where all digits assumed to be correct and $E_{3e}^{(fit)}$ from the fit (\ref{e-fit_3}).
}
\begin{tabular}{|c|c|c|}
\hline\hline
Z          &\    $E_{3e}$    \ &\ $E_{3e}^{(fit)}$\ \\
\tableline
\ 2.16  \  &\  -3.47810826   \ &\  -3.47810790 \  \\
\ 2.10  \  &\  -3.25509127   \ &\  -3.25509091 \  \\
\ 2.075 \  &\  -3.16479824   \ &\  -3.1647979  \  \\
\ 2.02  \  &\  -2.97184      \ &\  -2.97184    \  \\
\tableline
\hline
\end{tabular}
\end{table}

As a conclusion, we state that based on extrapolation of highly accurate results for the ground state energy for $(Z,2e)$ and $(Z,3e)$ to the critical charge, the branch point occurs with the exponent 3/2. This is in agreement with a statement made in \cite{Stillinger:1974}-\cite{Zamastil:2010} for $(Z,2e)$. Furthermore, it is in agreement with recent results for one-two electron molecular systems $(2Z,e), (3Z,e), (4Z,e)$ and $(2Z,2e), (3Z,2e)$ \cite{Turbiner:2011}. All these results indicate a universal nature of singularity at critical charge for Coulomb systems. So far, present authors are unable to give a physics explanation of this phenomenon. For all studied systems we did not see any indication of charge quantization.

\bigskip

\textit{\small Acknowledgements}. The authors are grateful to K. Pachucki for providing
the computer codes for making highly-accurate calculations of the ground state energy
of 2- and 3-electron atomic-type systems and extended explanations. A.V.T. wants to express the gratitude to B. Simon for useful remark. The research is supported in part by DGAPA grant IN115709 (Mexico). A.V.T. thanks the University Program FENOMEC (UNAM, Mexico)
for partial support.

\bigskip

\textit{Note Added}. Following a suggestion of one of the referees we made a preliminary numerical study of energy behavior {\it vs} $Z$ of the $2^1S$ state of two-electron system in domain $Z \in [1, 2]$. These calculations were carried out using a computer code kindly provided by Prof.~K.~Pachucki. We found the critical charge
$Z_{cr,2e}^{(2^1S)}\ =\ 1.02$
(cf.(\ref{ZcrHe3})). Energy curves in $\xi$ did not display in a clear way a behavior indicating quasi-crossings, see Fig. 1. It might be considered as a signal that their square-root branch points are situated far away from the real $\xi$ axis. A localization 
of these branch points requires a separate study which might be done elsewhere. 
A general situation with analytic structure of $E(\xi)$ remains unclear.

\begin{figure}[tb]
\begin{center}
   \includegraphics*[width=4in,angle=-90.0]{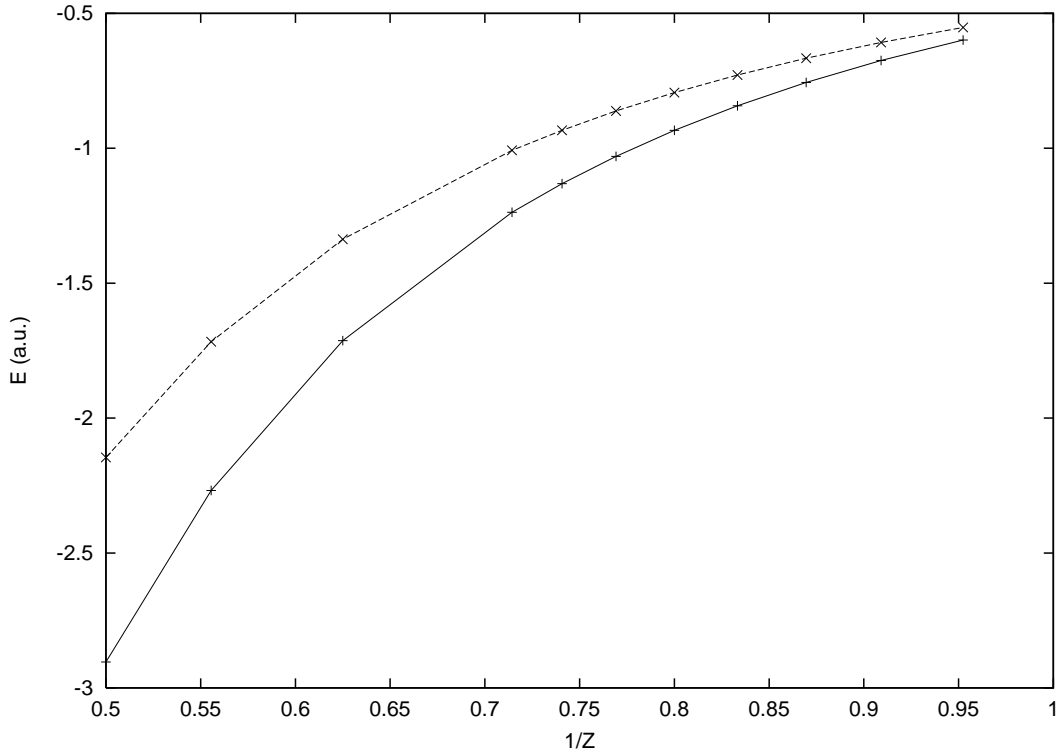}
    \caption{\label{figx} Energy Curves vs $Z$ for the $(Z,e,e)$ system for the spin-singlet $1^1S$ (solid line) and $2^1S$ (dash line) states}
\end{center}
\end{figure}

\end{document}